# A birds-eye (re)view of acid-suppression drugs, COVID-19, and the highly variable literature

Cameron Mura[1*], Saskia Preissner[2], Robert Preissner[3], Philip E. Bourne[1*]

### Author affiliations & correspondence

[1] School of Data Science and Department of Biomedical Engineering, University of Virginia, Charlottesville, VA; USA

[2] Department Oral and Maxillofacial Surgery, Charité–Universitätsmedizin Berlin, corporate member of Freie Universität Berlin, Humboldt-Universität zu Berlin, and Berlin Institute of Health, Augustenburger Platz 1, 13353 Berlin, Germany

[3] Institute of Physiology and Science-IT, Charité–Universitätsmedizin Berlin, corporate member of Freie Universität Berlin, Humboldt-Universität zu Berlin, and Berlin Institute of Health, Philippstrasse 12, 10115 Berlin, Germany

[*] Correspondence can be addressed to CM (cmura@virginia.edu) or PEB (peb6a@virginia.edu).

### Author contributions

Conception and design: CM
Acquisition, analysis and interpretation of data: CM, SP, RP, PEB
Drafting of the manuscript: CM, SP, RP, PEB
Supervision: RP, PEB
Critical revision of the manuscript: CM, SP, RP, PEB
Final approval: all authors

**Conflicts of interest**: No author declares a conflict of interest.

### Document information

| | |
|---|---|
| Last modified: | 23 April 2021 |
| Running title: | *Famotidine, COVID-19 & the Literature* |
| Keywords: | COVID-19; famotidine; histamine antagonist; knowledge graph; disease ontology |
| Abbreviations: | COD, context of disease; COVID-19, coronavirus disease 2019; DOB, degree of benefit; GPCR, G-protein coupled receptor; KG, knowledge graph; MOA, mechanism/mode of action; RDF, resource description framework; SARS-CoV-2, severe acute respiratory syndrome coronavirus 2 |
| Additional notes: | The main text includes 1 figure. |

This Note examines the evolution of a recent surge of information regarding the potential benefits of acid-suppression drugs in the context of the COVID-19 pandemic[1], with a particular eye on the great variability (and, thus, confusion) that has arisen across the reported findings, at least as regards the popular antacid famotidine. The degree of inconsistency and discordance reflects contradictory conclusions from independent clinical-based studies that took roughly similar approaches, in terms of both experimental design (retrospective, observational, cohort-based, etc.) and statistical analysis workflows (propensity-score matching and stratification into sub-cohorts, etc.). The contradictions and potential confusion have ramifications for clinicians faced with choosing therapeutically optimal courses of intervention: e.g., do potential benefits of famotidine suggest its use in a particular COVID-19 case? (If so, what administration route (oral, intravenous), dosage regimen, duration, etc. are likely optimal?) As succinctly put by Freedberg et al. this March[2], "*…several retrospective studies show relationships between famotidine and outcomes in COVID-19 and several do not.*" Beyond the pressing issue of potential therapeutic indications, the conflicting data and conclusions related to famotidine must be resolved before its inclusion/integration in ontological and knowledge graph (KG)–based frameworks, which in turn are useful for drug discovery and repurposing. As a broader methodological issue, note that reconciling inconsistencies would bolster the validity of meta-analyses which draw upon the relevant data-sources. And, perhaps most broadly, developing a system for treating inconsistencies would stand to improve the qualities of both (i) real world evidence-based studies (retrospective), on the one hand, and (ii) placebo-controlled, randomized multi-center clinical trials (prospective), on the other hand. In other words, bringing the two types of studies into consistency via some systematic approach would inherently improve the quality and utility of each type of study individually.

As a first step to begin systematically structuring the rapidly accumulating information—in the hopes of clarifying and perhaps reconciling the discrepancies, and eventually maturing the information into *clinically-actionable knowledge & understanding*—let us view this topic along three 'axes' implied by our opening sentence: namely, we consider *(i)* a *context-of-disease* (**COD**) axis, *(ii)* a *degree-of-[therapeutic]-benefit* (**DOB**) axis, and *(iii)* a *mechanism-of-action* (**MOA**) axis. We now treat each of these in turn; note that these 'axes' are of *nominal* type (in terms of classification levels and typologies[3]).

The ***MOA axis*** may be the most straightforward to conceptualize, as it simply describes the MOA of a putative drug—i.e., the mechanistic, molecular-level etiological basis (to the extent known), in whatever might be the most salient physiological pathways for that compound. As a concrete example, known gastroenterological acid-suppression agents may act either: (i) as proton-pump inhibitors (PPIs), sterically occluding proton efflux via the $H^+/K^+$-ATPase pumps that mediate the final step of acid release in the gastric mucosa, or (ii) as histamine type-2 receptor antagonists (H2RAs), whereby specific binding to this subtype of G-protein coupled receptor blocks the downstream signaling and effector cascades that otherwise would have been triggered by the cognate ligand (i.e., histamine)[4,5]. An example of a widely used PPI is omeprazole (e.g., Prilosec®), and famotidine (e.g., Pepcid®) and ranitidine (e.g., Zantac®) are examples of popular over-the-counter H2RAs. In the context of our present example, the MOAs of these two particular classes of GI drugs are mutually exclusive—i.e., a drug $\mathcal{D}$ acts by one pathway $\mathcal{P}_1$ (PPI-blocker) or another $\mathcal{P}_2$ (H2RA), but not both. Such basic, molecular-level selectiveness is not always the case, and, indeed, the approach of polypharmacology[6] relaxes the rigidity of the "*one gene, one disease, one drug*" view by recognizing that, in vivo, a given drug compound likely has multiple targets in multiple physiological pathways; indeed, generally speaking the potential multiplicity of drug⋯target linkages can be leveraged to beneficial effect. A careful consideration of 'off-target' effects (beneficial or detrimental) is beyond the scope of our present treatment. Here, we only note that a recent study[7] has provided a detailed and insightful analysis of possible on- and off-target properties of famotidine, particularly as regards (i) hypothesis testing of its MOA, and (ii) discrepancies that can arise among different studies because of differences in dosage, single- and multi-agent treatment regimes, pharmacokinetic-related properties (e.g., ADME), and so on; notably, that work concluded that off-target (i.e., non-H2R) pathways likely do not play a major role in the case of famotidine as a potential COVID-19 therapy.

One can envision developing more quantitative descriptors of the entities that populate the MOA axis and their interrelationships (to move from our nominal scale towards something more like an *ordinal scale*) by employing approaches like the *Coronavirus Infectious Disease Ontology*[8] (one of the many biomedical ontologies gathered at *BioPortal*[9]) to represent the targets of $\mathcal{D}$ in the $\mathcal{P}_1$, $\mathcal{P}_2$, etc. pathways, the possible side-effects, and so on. A potential source of confusion in the recent famotidine literature, taken in totality, is that recent meta-analyses[10-12] of retrospective, observational, cohort-based studies have not always enforced a clean delineation between the two MOAs mentioned above (i.e., PPI and H2RA), at least when drawing their conclusions. As a case in point, some primary studies have found that famotidine is beneficial whereas PPIs offered "no protective effect"[13]; another study found no positive effect of famotidine and also no deleterious effect of PPIs[14]; another found no association between PPIs and a different H2RA (ranitidine), in terms of likelihoods of both COVID-19 infection and death[15]; a pair of studies found no evidence for additional risks or benefits, for neither famotidine nor PPIs[16,17]; and, finally, another recent study found greater risks of association with morbidity or severe illness for famotidine and PPIs (each one, individually)[18]. Such incongruities, viewed holistically across all of the studies, result in internal inconsistencies when meta-analyses then draw upon a set of such studies: there is no clear 'combining' rule to harmonize otherwise inconsistent data sources and data values. Also, in some sense, comparing drugs with different MOAs, under the umbrella of a single, global analysis (be it a meta-analysis or a manual/human-expert analysis), is akin to comparing apples and oranges: they are of fundamentally different *types*, and attempts to compare them are *ill-posed* (in the ontological sense of semantic networks, structured reasoning, knowledge graphs, and so forth [reviewed in ref 19]). From a pharmacogenomics and drug discovery perspective, a key goal would be the development of ontology-based knowledge representation graphs (or *semantic models*) of the cellular pathways wherein famotidine might intervene as a drug (Figure 1). Such an approach could enable systematic, automated discovery (via *reasoning* over the graph) of potential new targets, new drug leads, and prediction of new drug/target pairs; however, constructing such a framework would require resolution of inconsistencies among the primary data used to build the semantic network's relationships (Fig 1A, and discussed below).

A few studies have begun considering the possible sources of discrepancies and heterogeneities, to which at least some of the inconsistencies in the famotidine/COVID-19 literature can be attributed. Some such sources relate to differences in study designs, or are otherwise methodological in nature—e.g. Freedberg et al.'s description of potential biases from residual confounders in the baseline characteristics of case-matched cohorts[20]. There is also the possibility of spurious links (or, inversely, masked associations) because of underlying physiological factors, e.g. Sethia et al.'s description[21] of the impacts of (i) potentially great differences in disease severity on treatment outcomes (without adequately accounting for such in case-matching and stratification methods to obtain sub-cohorts), (ii) heterogeneity in classification of the severity of illness, (iii) whether these variations are factored into the case-matching and stratification stages, (iv) variation in the regimen for famotidine treatment (administration route [oral, intravenous], dosage levels, timings with respect to onset of disease symptoms and duration of treatment)[22], and (v) confounding factors from co-medications or comorbidities among patients who do more/less well with famotidine. Some of the sources of differences will be at a more basic, molecular level, and likely quite difficult to elucidate: for instance, (i) Singh et al. note that fine-grained details such as the levels of calcium in various famotidine formulations may be "mechanistically relevant to disease outcomes"[23]; and (ii) in general, pharmacogenomic factors[24] will govern the potential efficacy of famotidine as a therapeutic, and these will vary on an individual, case-by-case basis.

The *degree-of-benefit* (*DOB*) axis is relatively easy to conceptualize: For famotidine (or in general any putative drug, $\mathcal{D}$), this categorical descriptor can be viewed as being essentially tripartite: "*neutral/no association*", "*pro-famotidine*", "*anti-famotidine*". The utility of the descriptor for a given drug compound will be specific to disease context—i.e., it would not hold universally across all diseases. Some drug $\mathcal{D}$ may be negatively indicated for a particular ailment or illness $\mathcal{I}_1$ (e.g., AIDS) whereas the same drug would be indicated in another illness $\mathcal{I}_2$ (e.g., COVID-19). That is to say, symbolically, that the DOB for a given drug in two different disease contexts may be dissimilar, $\mathcal{D}_{\text{DOB}}^{\text{AIDS}} \neq \mathcal{D}_{\text{DOB}}^{\text{COVID}}$. Such may hold, for example, with some PPIs: they have well-established records of efficacy in acid-suppression (gastroesophageal reflux disease, ulcers, etc.) and they may be valuable in seemingly unrelated contexts (e.g., lansoprazole as an inhibitor of rhinovirus infection[25]), yet are not necessarily beneficial in treating COVID-19. Idiosyncratic patterns such as these are drug-specific, pathway-dependent and disease-related, and they are especially salient in drug repurposing efforts (an endeavor that was recently reviewed for COVID-19[26], including in particular for GI drugs[27]; such efforts are facilitated by modern web-based resources, e.g. PROMISCUOUS[28]).

The *context-of-disease* (*COD*) axis accounts for the fact that, in addition to being subjective, the term 'beneficial' is also vague: it can be gauged by various *types* of outcomes/criteria, even for a single, given disease or set of indications. In the context COVID-19 and other infectious diseases, a putative drug may be 'beneficial' in terms of its impact upon (i) *transmissibility* (e.g., one study considered if usage of the H2RA ranitidine modulates susceptibility to SARS-CoV-2 infection[15]), (ii) *disease severity* indicators (e.g., likelihood of cases reaching the point of mechanical ventilation, the WHO Severity Index [Yeramaneni et al.[29] use this ordinal scale in COVID-19], or other measures), and (iii) *mortality rates*, frequencies or related metrics. Unlike the MOA axis, in our current case of COVID-19 and acid-suppressing drugs the entities along the COD axis can overlap—for example, there is no *a priori* reason why a given drug cannot be beneficial in two senses, e.g. by diminishing transmissibility of SARS-CoV-2 *and* by reducing the severity of the disease trajectory once it has been contracted (note, though, that in general different cellular/molecular-scale pathways [MOAs] can underlie similar organismal-level outcomes).

We suspect that some incongruencies in how these MOA, COD and DOB axes have been (implicitly) treated in the various, independent studies is what has led to the contradictory indications regarding famotidine and COVID-19 (*anti-famotidine*, *neutral*, *pro-famotidine*) in the published literature. Harmonizing the findings across the literature is an important goal from the perspectives of both bioinformatics and clinical standards-of-care/best practices for a given disease.

Finally, now to trace the evolution of what we know about the potential benefits (*DOB*) of H2RAs, PPIs, etc. (different *MOAs*) on the transmission, severity, and mortality (various *CODs*) of COVID-19, let us consider the many literature reports that have accumulated in the past year.  By grouping these studies conceptually and thematically, we can begin to identify the following three 'Eras' in the progression of the literature (and our understanding) as regards famotidine and COVID-19:

- *Era 1*: *Primary research studies by independent groups*: These ≈10 analyses generally have been *retrospective*, *observational* (some *cohort-based*, some *case-series*) and *single-site/center*. Most of the studies attempted to *statistically adjust for confounders* (e.g., via propensity score matching), though with varying degrees of rigor and caution (see the study cited in *Era 3*, below).  Despite many similarities in design, there are two contradictory sets of studies: (i) a handful of studies concluded in favor of using famotidine in COVID-19 (Freedberg et al.[13], Janowitz et al.[30], Mather et al.[31], Hogan et al.[32], and Sethia et al.[21]), while (ii) a roughly equal number did not[14-16, 18, 29].  In the second category of reports that were less enthusiastic about famotidine, some studies did indicate against the usage of famotidine (Cheung et al.[16], Yeramaneni et al.[29], and Zhou et al.[18]) while others found no association for famotidine or PPIs and COVID-19 (Fan et al.[15], Elmunzer et al.[14]).

- *Era 2*: *Meta-analyses of the primary literature reports*: Three such analyses have appeared thus far, with again varying results: Sethia et al.[11] were pro-famotidine, while Sun et al.[12] and Kamal et al.[10] were more neutral/negative, determining for the most part that no association (positive or negative) was statistically justifiable.  Why the inconsistency in whether or not famotidine is indicated for COVID-19, even at the level of a *meta*-analysis?  One contributing factor could be that different database inclusion criteria were used in these different meta-analyses (see the Methods sections in each of refs 10-12). For instance, Sun et al.[12] drew upon several Chinese databases and did not sample the medRxiv or SSRN preprint collections, whereas Sethia et al.[11] included medRix and SSRN but not the several Chinese databases. This could be a source of discrepancy between the meta-analyses.  Furthermore, it is perhaps unsurprising that the meta-analysis which did reach a more favorable conclusion as regards indicating famotidine (i.e., Sethia et al.) leaned most heavily on the five studies from the primary literature which were, themselves, most strongly in favor of famotidine on average (i.e., refs 13, 21, 30-32); in contrast, the meta-analysis which found no/less-favorable association between famotidine and positive COVID outcomes (i.e., Sun et al.) included a subset of case/cohort-studies that generally reached less favorable conclusions (e.g., refs 16, 17, 29). The third meta-analysis, of Kamal et al., does not serve as a 'tie-breaker' here, as it finds a "lack of consistent association" between COVID-19 outcomes and the use of acid-suppression drugs (for both famotidine and PPIs). Finally, as can generally occur in bibliometric meta-analyses, a 'positive-outcome' publication bias[33] may exist, with negative findings never having made it to the primary, peer-reviewed literature.

- *Era 3*: *Origins of the discrepancies?*: Without resorting to a *meta-meta-analysis*(!), Etminan et al.[34] recently supplied a thorough and incisive critique of factors that may limit the consistency of the conclusions drawn thus far from the primarily retrospective, observational, single-site studies (such as those in refs 13, 14, 31).  In particular, likely sources were identified for several types of biases, including residual *confounding bias* and *sparse-data bias*, *immortal time bias* (and somewhat related *selection bias* effects), and reverse causality bias (the latter can be understood via *causal directed acyclic graph* [cDAG] models, which can elucidate the structure of these epidemiological biases).  Here, we simple note that: (i) Some of these sources of bias are rather more difficult to control for,

*Famotidine, COVID-19 & the Literature*or statistically identify and account for, than others (because of causal relationships, limited data, etc.); (ii) In principle, biases, in and of themselves, do not necessarily result in inconsistencies among a series of primary research studies. That is, *bias* and *inconsistency* are not equivalent phenomena: conceivably, each study within a series of studies could suffer from similar biases, and nevertheless still be self-consistent. Regardless, Etminan et al. highlight several improvements that can be made in future approaches to statistically elucidate famotidine ↔ COVID-19 relationships.

Perhaps some inconsistencies can be resolved by viewing famotidine and its potential roles in COVID-19 through a more mechanistic, molecular-level lens? Alongside the observational, patient-based studies, the potential MOA for a therapeutic role of famotidine has been explored in several publications: Ennis & Tiligada's recent review[5] offers a cogent and authoritative treatment of the connection between histamine receptor antagonists (e.g. famotidine) and COVID-19, while histamine release theory was also the subject of Eldanasory et al.[4] and Ghosh et al.'s[35] accounts of the role of this versatile signaling molecule in pro-inflammatory pathways—including the destructive 'cytokine storm' that appears to underlie much of the pathophysiology of COVID-19, at least with respect to pulmonary consequences (fibrosis, etc.) and the resultant acute respiratory distress. More generally, famotidine and other GI-related drugs appear in Tarighi et al.'s comprehensive review[26] of drug repurposing in the age of COVID, and Singh et al.'s recent letter[23] considers the question of what (molecularly) underlies the efficacy of treatment of COVID with famotidine? These mechanistic, biomolecular directions offer hope, as it is often the case that multiple disparate (and seemingly contradictory) phenomena, or sets of observations, at a macroscopic/organismal level become reconciled when viewed at the molecular level, in terms of the underlying cellular and physiological pathways. Indeed, we suspect that the reconciliation and harmonization of independent, contradictory findings—such as for famotidine and COVID-19—ultimately will lie along two paths: (i) molecular-level examination of whatever pathways are thought to be most salient for given macroscopic phenomena (clinical-level observations), and (ii) multi-scale, integrative modeling of the interaction and build-up of microscopic phenomena (e.g., drug-protein interactions) through many hierarchical levels (organellar ↣ subcellular ↣ cellular ↣ tissues ↣ ⋯) up through to the organismal level, and eventually even populations (therapeutic outcomes, like altering the course of an infection).

Ontologies, and the way forward?: A key motivation for reconciling conflicts in the famotidine × COVID-19 literature, in terms of basic drug discovery/repurposing and modeling COVID-19-drug interactions, stems from the importance of data consistency and harmonization in constructing ontology-based knowledge graphs of biomolecular systems. Figure 1 shows a diagrammatic representation of the types of *Resource Description Framework* (RDF) 'triples' that represent the relationships between distinct entities (refs [36-38] offer introductions/primers on ontologies). That example also denotes other histamine receptors (H1R, H3R, H4R) in order to emphasize that any given pairwise relationship does not exist in isolation: all simpler relationships are embedded within physiological contexts that typically involve many entities (with varying degrees of confidence in the linkages). While RDF triples capture relationships between entities, the other aspect of an ontology is a *structuring* or *organization* (typically hierarchical) for the collection of classes/concepts (of which a given entity is an *instance*) that span the ontology's domain of knowledge. A gene ontology (GO)[39] "ancestor chart", shown in Fig 1B, illustrates such a concept hierarchy for the term GO:0031808, corresponding to H2R-binding. The excerpt of an ontological map of host–coronavirus

Mura et al. (2021)    6/10

drug/interactions (Fig 1C) is a graph representation that reflects both aforementioned aspects of an ontology: namely, (i) the *structural interrelationships* among concepts/entities (some broad, some specific, some subsets of others, etc.), as well as (ii) the *pairwise associations* (RDF triples) from which such knowledge graphs are built. Such 'semantic network' approaches are compelling and powerful for many reasons (reviewed in refs 19, 36), including that the knowledge graph can be used for drug discovery, pharmacogenomics modeling, and 'big data' endeavors (data integration across scales); one manner in which this is achieved is by 'reasoning' over the KG, beyond the initial confines (or domain) of the data sources. The success of all such efforts rests upon the data being systematized against a controlled vocabulary, syntactically well-formed, and so on. A key determinant of our ability to build such graphs is that there not be internal inconsistencies, such as "$\checkmark \mathcal{F}amo_{\text{DOB}}^{\text{COVID}}$" and "$\times \mathcal{F}amo_{\text{DOB}}^{\text{COVID}}$". Resolving inconsistencies and inaccuracies in the data-sources can mitigate the 'percolation' of such errors through the graph, thereby limiting potentially erroneous downstream conclusions (such conclusions are often reached by applying machine learning approaches to perform statistical inference on the graph-based structures).

Together with the findings from several sets of prospective clinical trials that have been underway—including NCT04504240, NCT04370262 and NCT04545008 in the U.S., and other efforts internationally (e.g., ref 40)—we anticipate that detailed biomolecular studies can help clarify the contradictory relationships that have been reported thus far between famotidine and COVID-19. That, in turn, will enable the creation of more robust, efficacious and predictive ontological frameworks for drug discovery and repurposing.

## Acknowledgements
We thank RW Malone for alerting us to some recent studies in the literature, particularly as regards the potential mechanisms of action of famotidine.

## Funding
This work was supported by TRR295, KFO339, and DFG PR1562/1-1 (RP). Portions of this work were also supported by the University of Virginia School of Data Science and by NSF Career award MCB-1350957.

## Figures [1, total]

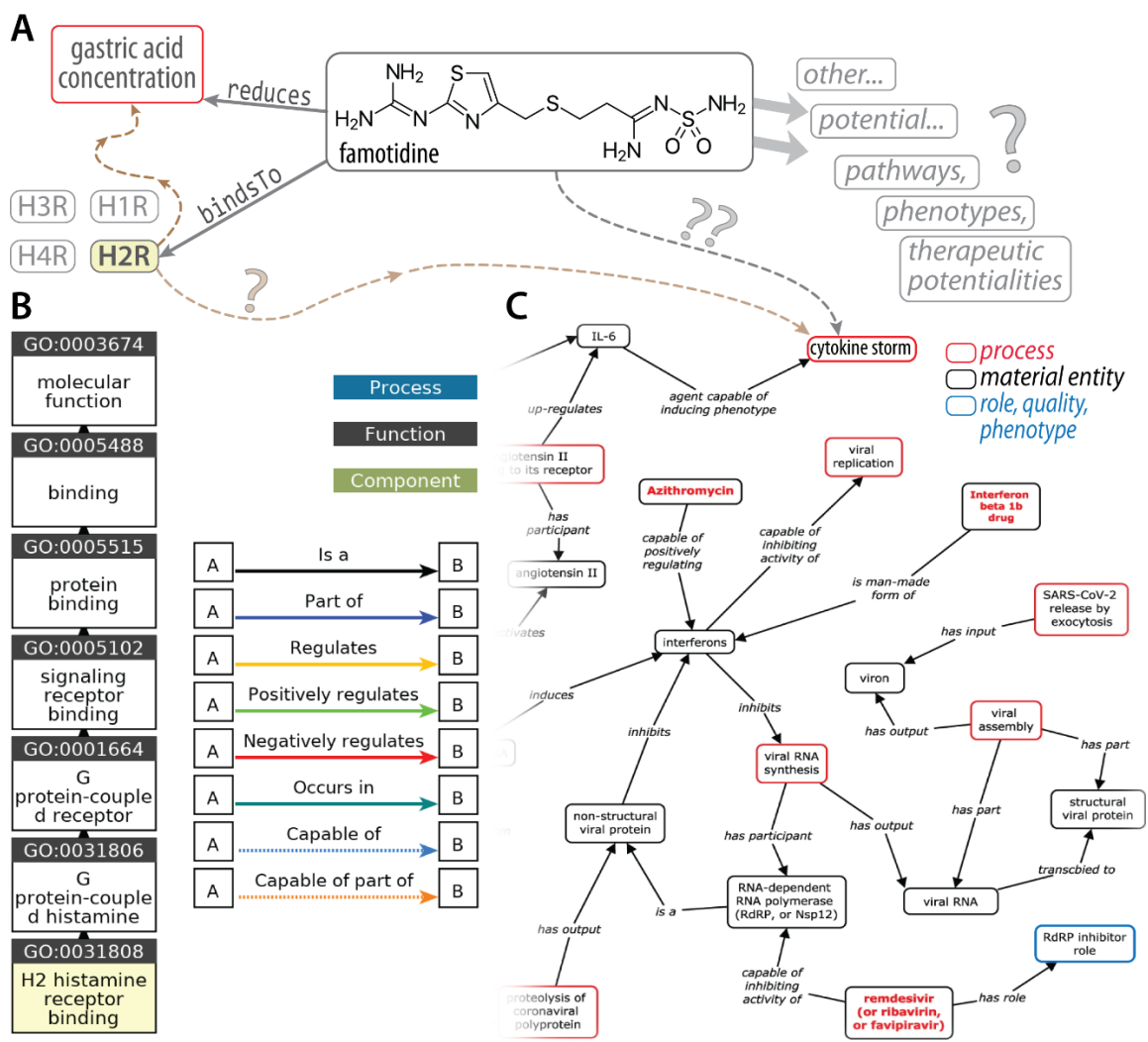

Figure 1. **An ontological perspective on famotidine and COVID-19.** This diagram (A) illustrates the types of RDF triple associations, consisting of a [`subject`, *predicate*, `object`] triplet, that express the relations between entities that underlie an ontology. In this simplified example, one such triple would be [`famotidine`, *bindsTo*, `H2R`] (histamine type-2 receptor) and another is [`famotidine`, *reduces*, `gastric acid concentration`]; H1R, H3R and H4R denote other subtypes of histamine receptors. The brown dashed arrows (at left) are placeholders to signify the H2R-triggered signaling cascades that ultimately modulate gastric acid levels. Question marks ('?') decorate linkages where our knowledge is either quite tenuous or vague. The GO "ancestor chart", in (B), illustrates a concept hierarchy for the term GO:0031808, corresponding to a particular *molecular function* named "`H2 histamine receptor binding`"; traversing this concept/class hierarchy, from the leaf (bottom) to the root (top; "`molecular function`"), corresponds to a traversal of successively broader categories of concepts/classes (e.g., the H2R `is_a` type of [i.e., a subset of] GPCRs). Panel (C), adapted from Liu et al.[8], is an excerpt of a KG of host–coronavirus drug/interactions, including at the upper-right a putative link between famotidine and the cytokine storm.

...